\documentclass[useAMS,usenatbib]{mn2e}
\usepackage{amssymb}
\usepackage{graphicx}

\title[]{Asteroseismology of DAV star EC14012-1446, mode identification and model fittings}
\author[Y. H. Chen, Y. Li]{Y. H. Chen$^{1,2,3}$\thanks{E-mail: yanhuichen1987@ynao.ac.cn, ly@ynao.ac.cn} and Y. Li$^{1,2}$\\
$^{1}$Yunnan Observatories, Chinese Academy of Sciences, Kunming 650011, China\\
$^{2}$Key Laboratory for the Structure and Evolution of Celestial Objects, Chinese Academy of Sciences, Kunming 650011, China\\
$^{3}$University of Chinese Academy of Sciences, Beijing 100049, China}

\begin{document}

\date{Accepted: }

\pagerange{\pageref{firstpage}--\pageref{lastpage}} \pubyear{????}

\maketitle

\label{firstpage}

\begin{abstract}
EC14012-1446 was observed by Handler et al. in April 2004, June 2004, May 2005, and April 2007, and by Provencal et al. in 2008. We review the observations together and obtain 34 independent frequencies. According to the frequency splitting and the asymptotic period spacing law, we identify 6 $l$=1 modes, 4 $l$=2 modes, 5 $l$=3 modes, 10 $l$=1 or 2 modes. Grids of white dwarf models are generated by WDEC with H, He, C, O diffusion in a four-parameter space. The core compositions are directly from white dwarf models generated by MESA. The best-fitting model has $M_{*}$=0.710\,$M_{\odot}$, $T_{\rm eff}$=12200\,K, log($M_{\rm He}/M_{*}$)=-2.5, log($M_{\rm H}/M_{*}$)=-7.0, log\,$g$=8.261, and $\phi$=3.185\,s. There are 4, 2, and 1 modes identified as trapped in H envelope for observed $l$=1, 2, and 3 modes, respectively. Trapped modes jump the queue of uniform period spacing.
\end{abstract}


\section{Introduction}

About 80\% of white dwarfs show DA spectral class. A DA white dwarf star consists of a hydrogen atmosphere covering an intermediate helium layer and a carbon/oxygen core. According to existence of partial hydrogen ionization and subsurface convection zone, a DA star will pulsate when its effective temperature ($T_{\rm eff}$) is between 10850\,K and 12270\,K (Castanheira et al. 2007). With buoyancy acting as the restoring force, DAV stars are pulsating in $g$-modes. Asteroseismology is an unique tool to detect the inner structure of DAV stars, which requires enough observed frequencies, reliable mode identifications, and realistic stellar models.

EC14012-1446, also called WD1401-147, was first identified as a DAV star by Stobie et al. (1995). They observed it for 4 nights, and then derived 5 independent frequencies from the obtained light curves. Subsequently, Handler et al. observed it several times in April and June 2004, May 2005, and April 2007, and obtained 19 independent frequencies (Handler et al. 2008). Recently, Provencal et al. observed EC14012-1446 again in the WET run XCOV26 in 2008 and obtained 19 independent frequencies (Provencal et al. 2012). On the other hand, Bergeron et al. (2004) obtained $T_{\rm eff}$=11900\,K, log\,$g$=8.16, and $M_{*}/M_{\odot}$=0.70 for EC14012-1446, when they studied the purity of DAV star instability strip on the basis of spectroscopy. High-resolution spectra of more than 1000 white dwarfs were obtained by ESO Supernova Ia Progenitor Survey (SPY) (Koester et al. 2009). The best atmospheric model for EC14012-1446 results in $T_{\rm eff}$=11768\,K$\pm$23\,K and log\,$g$=8.080$\pm$0.008.

Mode identification is an important task for asteroseismological studies. An eigenmode can be characterized by three indices ($k$, $l$, $m$), which are respectively the radial order, the spherical harmonic degree, and the azimuthal number. Frequencies of pulsation modes with definite spherical harmonic degrees play a role as tick marks on a ruler for theoretical modeling of pulsating white dwarfs. In particular, stellar rotation can split a pulsation frequency into several ones. The approximate formula between frequency splitting ($\delta\nu_{n,l}$) and rotational period ($P_{\rm rot}$) is derived by Brickhill (1975) as
\begin{equation}
m\delta\nu_{n,l}=\nu_{n,l,m}-\nu_{n,l,0}=\frac{m}{P_{\rm rot}}(1-\frac{1}{l(l+1)}),
\end{equation}
\noindent where $m$ can be taken from $-l$ to $l$, leading to totally 2$l$+1 different values. According to Eq.\,(1), modes with $l$=1 form a triplet and modes with $l$=2 form a quintuplet. Therefore, if triplets or quintuplets are derived from observations of a pulsating star, they can be reliably identified as rotational splitting of $l$=1 or $l$=2 modes, respectively.

For EC14012-1446, Stobie et al. (1995) did not find any rotational splitting phenomenon. Later, based on those independent frequencies they had obtained, Handler et al. (2008) declared the discovery of two triplets with an average frequency splitting of 9.55\,$\mu$Hz. However, Provencal et al. (2012) did not confirm the discovery of Handler et al. (2008). Instead, they identified a new triplet with an average frequency splitting of 3.79\,$\mu$Hz, which was in disagreement with result of Handler et al. (2008).

By comparing carefully the independent frequencies obtained by Handler et al. (2008) and by Provencal et al. (2012), we have noticed that two frequencies of a triplet (around 1633\,$\mu$Hz and 1623\,$\mu$Hz) in the result of Handler et al. (2008) are also present in the result of Provencal et al. (2012). Consequently, we suggest that the frequency splitting due to stellar rotation is around 9.55\,$\mu$Hz for EC14012-1446. Based on this argument, we propose in the present paper a new scheme of mode identification for EC14012-1446, and accordingly make new theoretical models to fit the observed pulsation frequencies.

In Sect. 2, we propose our new scheme of mode identification in order to solve the frequency splitting problem between results of Handler et al. (2008) and Provencal et al. (2012). In Sect. 3, we try to do model fittings for EC14012-1446 based on our new result of mode identification. Input physics and model calculations are described in Sect. 3.1, and selection of the best-fitting model is discussed in Sect. 3.2. Mode trapping effect is discussed in Sect. 3.3. In Sect. 4, we compare fitting results of the best-fitting model with other spectroscopy and asteroseismology studies. In Sect. 5, we summarize our conclusions and make some discussions of our results.

\section{Mode identification for EC14012-1446}

\begin{table*}
\begin{center}
\begin{tabular}{lccccccccccccccc}
\hline
&&&\multicolumn{2}{c}{April 2004}&\multicolumn{2}{c}{June 2004}&\multicolumn{2}{c}{May 2005}&\multicolumn{2}{c}{April2007}&\multicolumn{2}{c}{2008}\\
\hline
ID      &Freq.    &Peri.   & Freq.   &Ampl.    & Freq.   &Ampl.    & Freq.   &Ampl.    & Freq.   &Ampl.    & Freq.   &Ampl.    \\
        &($\mu$Hz)&(s)     &($\mu$Hz)&(mma)    &($\mu$Hz)&(mma)    &($\mu$Hz)&(mma)    &($\mu$Hz)&(mma)    &($\mu$Hz)&(mma)    \\
\hline
$f_{1}$ &2856.155 &350.121 &         &         &         &         &         &         &         &         &2856.155 &$\,$ 2.0 \\
$f_{2}$ &2508.060 &398.715 &         &         &         &         &         &         &         &         &2508.060 &$\,$ 2.1 \\
$f_{3}$ &2504.871 &399.222 &2504.86  &$\,$ 8.7 &2504.98  &$\,$ 8.1 &2504.97  &$\,$ 6.8 &2504.65  &$\,$ 8.6 &2504.897 &$\,$12.7 \\
$f_{4}$ &2304.745 &433.887 &         &         &         &         &         &         &         &         &2304.745 &$\,$ 4.7 \\
$f_{5}$ &1891.142 &528.781 &         &         &         &         &         &         &         &         &1891.142 &$\,$ 3.8 \\
$f_{6}$ &1887.519 &529.796 &1887.47  &$\,$ 8.9 &1887.79  &$\,$ 9.3 &1887.34  &$\,$12.3 &1887.59  &$\,$15.2 &1887.404 &$\,$20.7 \\
$f_{7}$ &1883.555 &530.911 &         &         &         &         &         &         &         &         &1883.555 &$\,$ 1.5 \\
$f_{8}$ &1860.248 &537.563 &         &         &         &         &         &         &         &         &1860.248 &$\,$ 6.4 \\
$f_{9}$ &1774.989 &563.384 &         &         &         &         &         &         &         &         &1774.989 &$\,$ 7.2 \\
$f_{10}$&1643.368 &608.506 &1643.40  &$\,$14.2 &1642.96  &$\,$13.6 &1643.03  &$\,$ 7.0 &1644.08  &$\,$ 4.1 &         &         \\
$f_{11}$&1633.653 &612.125 &1633.36  &$\,$48.1 &1633.60  &$\,$48.3 &1633.71  &$\,$33.5 &1633.69  &$\,$ 4.7 &1633.907 &$\,$25.7 \\
$f_{12}$&1623.573 &615.925 &1623.28  &$\,$11.2 &1623.51  &$\,$ 9.0 &1623.86  &$\,$10.5 &1623.20  &$\,$18.6 &1624.015 &$\,$ 3.1 \\
$f_{13}$&1548.146 &645.933 &         &         &         &         &         &         &         &         &1548.146 &$\,$ 7.9 \\
$f_{14}$&1521.575 &657.214 &         &         &         &         &         &         &         &         &1521.575 &$\,$ 2.2 \\
$f_{15}$&1484.130 &673.795 &1484.13  &$\,$ 2.6 &         &         &         &         &         &         &         &         \\
$f_{16}$&1473.783 &678.526 &1474.12  &$\,$ 9.3 &1474.95  &$\,$ 9.2 &1473.02  &$\,$ 5.6 &1473.04  &$\,$ 6.3 &         &         \\
$f_{17}$&1464.097 &683.015 &1464.17  &$\,$ 3.2 &1464.17  &$\,$ 6.8 &         &         &1463.95  &$\,$ 6.4 &         &         \\
$f_{18}$&1418.369 &705.035 &         &         &         &         &         &         &         &         &1418.369 &$\,$ 1.2 \\
$f_{19}$&1399.065 &714.763 &         &         &         &         &1398.29  &$\,$10.9 &1399.84  &$\,$13.6 &         &         \\
$f_{20}$&1394.910 &716.892 &         &         &         &         &1394.91  &$\,$ 7.3 &         &         &         &         \\
$f_{21}$&1385.320 &721.855 &1385.50  &$\,$24.4 &1385.62  &$\,$35.1 &1384.84  &$\,$40.6 &1385.32  &$\,$44.1 &         &         \\
$f_{22}$&1381.970 &723.605 &         &         &         &         &1381.97  &$\,$ 9.6 &         &         &         &         \\
$f_{23}$&1375.383 &727.070 &1375.53  &$\,$ 5.5 &1375.02  &$\,$ 6.9 &1375.60  &$\,$ 4.7 &         &         &         &         \\
$f_{24}$&1371.390 &729.187 &         &         &         &         &1371.89  &$\,$ 9.1 &1370.89  &$\,$20.0 &         &         \\
$f_{25}$&1299.810 &769.343 &         &         &         &         &1299.00  &$\,$39.8 &1300.62  &$\,$63.6 &         &         \\
$f_{26}$&1295.730 &771.766 &         &         &         &         &1295.73  &$\,$ 8.7 &         &         &         &         \\
$f_{27}$&1289.280 &775.627 &         &         &         &         &1289.28  &$\,$ 5.3 &         &         &         &         \\
$f_{28}$&1241.403 &805.540 &         &         &         &         &         &         &         &         &1241.403 &$\,$ 1.2 \\
$f_{29}$&1155.925 &865.108 &         &         &         &         &         &         &         &         &1155.925 &$\,$ 1.9 \\
$f_{30}$&1132.890 &882.698 &1132.89  &$\,$ 2.9 &         &         &         &         &         &         &         &         \\
$f_{31}$&1104.252 &905.591 &         &         &         &         &         &         &         &         &1104.252 &$\,$ 2.2 \\
$f_{32}$&1021.139 &979.299 &         &         &         &         &         &         &         &         &1021.139 &$\,$ 1.7 \\
$f_{33}$&935.380  &1069.085&         &         &         &         &         &         &         &         &935.380  &$\,$ 2.7 \\
$f_{34}$&821.390  &1217.448&821.26   &$\,$ 7.1 &821.52   &$\,$ 7.8 &         &         &         &         &         &         \\
\hline
\end{tabular}
\caption{The determined frequency. The first column shows frequency ID taking the observations of Handler et al. (2008) and Provencal et al. (2012) into account. Freq. = frequency in $\mu$Hz, Peri. = period in seconds, and Ampl. = amplitude in mmag.}
\end{center}
\end{table*}

The independent frequencies obtained by Handler et al. (2008) and Provencal et al. (2012) are listed in Table 1. The first column shows frequency ID. The second column shows independent frequency (Freq.), which is an averaged value of results derived from observations in April 2004, June 2004, May 2005, April 2007, and 2008. The third column is the corresponding pulsation period (Peri.). Frequencies and amplitudes obtained in each observation are then shown in the following columns. There are 34 independent frequencies altogether.

 \subsection{Frequency splitting due to stellar rotation}

Handler et al. (2008) identified $f_{10}$, $f_{11}$, and $f_{12}$ as a triplet and $f_{15}$, $f_{16}$, and $f_{17}$ as another one, and determined an averaged frequency splitting to be 9.55\,$\mu$Hz. It can be noticed in Table 1 that these two triplets appeared simultaneously in April 2004. Later, the first triplet still appeared in June 2004, May 2005, and April 2007, while only parts of the second triplet appeared again. The absence of a particular mode in a triplet might be due to too small its amplitude to be observable.

Provencal et al. (2012) did not find both above triplets in result of their observations. Instead, they identified $f_{5}$, $f_{6}$, and $f_{7}$ as a new triplet with an averaged frequency splitting of 3.79\,$\mu$Hz. It is evident that their result on the frequency splitting is in disagreement with that of Handler et al. (2008). However, Provencal et al. (2012) suggested that $f_{11}$ and $f_{12}$ might also be a doublet.

It can be noticed in Table 1, however, that $f_{11}$ and $f_{12}$ appeared repeatedly in results of both Handler et al. (2008) and Provencal et al. (2012). Mode $f_{10}$ did not appear in result of observations in 2008, possibly due to its too small amplitude during that time. According to this argument, we suggest that $f_{10}$, $f_{11}$, and $f_{12}$ form a triplet with an approximate frequency splitting of about 10\,$\mu$Hz, just as originally suggested by Handler et al. (2008). We also suggest that $f_{15}$, $f_{16}$, and $f_{17}$ form another triplet, the same as Handler et al. (2008) did.

Based on above considerations, we suggest that $f_{20}$, $f_{21}$, and $f_{23}$ can also be identified as a triplet. It can be noticed in Table 1 that all of the three components appeared in result of observations in May 2005. In addition, $f_{21}$ and $f_{23}$ also appeared in results of both observations in 2004, while $f_{21}$ appeared again in result of observations in April 2007.

Furthermore, $f_{25}$ and $f_{27}$ can be identified as a doublet with a frequency splitting of 9.72\,$\mu$Hz. However, it is difficult to determine which one is the mode of $m$=0. We notice in Table 1, that the amplitude of $f_{11}$ (the central mode of the triplet) is usually larger than amplitudes of $f_{10}$ and $f_{12}$, except for result of observations in April 2007. Similarly, the amplitude of $f_{16}$ is usually larger than amplitudes of $f_{15}$ and $f_{17}$, except for result of observations in April 2007. In addition, the amplitude of $f_{21}$ is always the largest among the three components. Therefore, we may hypothesize that the mode of $m$=0 usually has the largest amplitude among all modes in the same multiplet. According to result of observations in May 2005, the amplitude of $f_{25}$ is much larger than that of $f_{27}$. We thus suggest that $f_{25}$ is the $m$=0 mode.

According to Eq.\,(1), the frequency splitting of $l$=2 modes due to stellar rotation is related to that of $l$=1 modes by (Winget et al. 1991):
\begin{equation}
\frac{\delta\nu_{n,1}}{\delta\nu_{n,2}}=\frac{3}{5}.
\end{equation}
\noindent Taking 10\,$\mu$Hz as the approximated frequency splitting of $l$=1 modes, we estimate the frequency splitting of $l$=2 modes to be about 16\,$\mu$Hz. It can be found in Table 1 that $f_{19}$ and $f_{22}$, which result in a frequency difference of 16.32\,$\mu$Hz in result of May 2005, may be identified as two neighbouring components of a quintuplet. The amplitude of $f_{19}$ is larger than that of $f_{22}$ and $f_{19}$ is also appeared in result of April 2007. Therefore, we identify $f_{19}$ as the mode of $m=0$. In addition, $f_{5}$ and $f_{8}$, with a frequency splitting of 30.89\,$\mu$Hz in result of 2008, may also be identified as two components with $\Delta$m=2 of another quintuplet. It can be seen in Table 1 that the amplitude of $f_{5}$ is about a half of that of $f_{8}$. Then, $f_{8}$ may be identified as the mode of $m=0$.

Handler et al. (2008) regarded frequencies around $f_{21}$ as complicated structure and hypothesized that they could be members of multiplets with different $l$ values. We solve this problem by identifying $f_{20}$, $f_{21}$, and $f_{23}$ as a triplet, meanwhile identifying $f_{19}$ and $f_{22}$ as two components of a quintuplet.

We list in Table 2 the identified rotational splitting of eigenmodes with azimuthal numbers on EC14012-1446. Based on the identified triplets and quintuplets, we estimate according to Eq.\,(1) a rotation period of about 0.58\,d for EC14012-1446, which is basically the same as 0.61\,d given by Handler et al. (2008).

\begin{table}
\begin{center}
\begin{tabular}{lccccccccccccccc}
\hline
ID      &Freq.    &$\delta$F&$l$   &$m$         \\
        &($\mu$Hz)&($\mu$Hz)&      &            \\
\hline
$f_{5}$ &1891.142 &         &2     &$\,$$+2$?   \\
        &         &30.894   &      &            \\
$f_{8}$ &1860.248 &         &2     &$\,$$0$?    \\
        &         &         &      &            \\
        &         &         &      &            \\
$f_{10}$&1643.368 &         &1     &$\,$$+1$    \\
        &         &9.715    &      &            \\
$f_{11}$&1633.653 &         &1     &$\,$  0     \\
        &         &10.080   &      &            \\
$f_{12}$&1623.573 &         &1     &$\,$$-1$    \\
        &         &         &      &            \\
        &         &         &      &            \\
$f_{15}$&1484.130 &         &1     &$\,$$+1$    \\
        &         &10.347   &      &            \\
$f_{16}$&1473.783 &         &1     &$\,$  0     \\
        &         &9.686    &      &            \\
$f_{17}$&1464.097 &         &1     &$\,$$-1$    \\
        &         &         &      &            \\
        &         &         &      &            \\
$f_{19}$&1399.065 &         &2     &$\,$  0?    \\
        &         &17.095   &      &            \\
$f_{22}$&1381.970 &         &2     &$\,$$-1$?   \\
        &         &         &      &            \\
        &         &         &      &            \\
$f_{20}$&1394.910 &         &1     &$\,$$+1$    \\
        &         &9.590    &      &            \\
$f_{21}$&1385.320 &         &1     &$\,$  0     \\
        &         &9.937    &      &            \\
$f_{23}$&1375.383 &         &1     &$\,$$-1$    \\
        &         &         &      &            \\
        &         &         &      &            \\
$f_{25}$&1299.810 &         &1     &$\,$  0?    \\
        &         &10.530   &      &            \\
$f_{27}$&1289.280 &         &1     &$\,$$-1$?   \\
\hline
\end{tabular}
\caption{Possible frequency splitting. $\delta$F = frequency separation in $\mu$Hz.}
\end{center}
\end{table}

 \subsection{Analysis according to the asymptotic theory}

\begin{table}
\begin{center}
\begin{tabular}{lccccccccc}
\hline
ID       &Freq.    &Peri.   &Peri. Diff.&$l$       \\
         &($\mu$Hz)&(s)     &(s)        &          \\
\hline
$f_{3}$  &2504.871 &399.222 &130.574    &$\,$1?    \\
$f_{6}$  &1887.519 &529.796 &82.329     &$\,$1?    \\
$f_{11}$ &1633.653 &612.125 &66.401     &$\,$1     \\
$f_{16}$ &1473.783 &678.526 &43.329     &$\,$1     \\
$f_{21}$ &1385.320 &721.855 &47.488     &$\,$1     \\
$f_{25}$ &1299.810 &769.343 &           &$\,$1     \\
         &         &        &           &          \\
$f_{2}$  &2508.060 &398.715 &138.848    &$\,$2?    \\
$f_{8}$  &1860.248 &537.563 &177.200    &$\,$2     \\
$f_{19}$ &1399.065 &714.763 &57.003     &$\,$2     \\
$f_{26}$ &1295.730 &771.766 &           &$\,$2?    \\
         &         &        &           &          \\
$f_{7}$  &1883.555 &530.911 &174.124    &$\,$3?    \\
$f_{18}$ &1418.369 &705.035 &24.152     &$\,$3?    \\
$f_{24}$ &1371.390 &729.187 &           &$\,$3?    \\
         &         &        &           &          \\
$f_{1}$  &2856.155 &350.121 &           &$\,$1 or 2\\
$f_{4}$  &2304.745 &433.887 &           &$\,$1 or 2\\
$f_{9}$  &1774.989 &563.384 &           &$\,$1 or 2\\
$f_{13}$ &1548.146 &645.933 &           &$\,$1 or 2\\
$f_{14}$ &1521.575 &657.214 &           &$\,$1 or 2\\
$f_{28}$ &1241.403 &805.540 &           &$\,$1 or 2\\
$f_{29}$ &1155.925 &865.108 &17.590     &$\,$3?    \\
$f_{30}$ &1132.890 &882.698 &           &$\,$3?    \\
$f_{31}$ &1104.252 &905.591 &           &$\,$1 or 2\\
$f_{32}$ &1021.139 &979.299 &           &$\,$1 or 2\\
$f_{33}$ &935.380  &1069.085&           &$\,$1 or 2\\
$f_{34}$ &821.390  &1217.448&           &$\,$1 or 2\\
\hline
\end{tabular}
\caption{Mode identification. Peri. Diff. shows the period difference between two modes in seconds.}
\end{center}
\end{table}

According to the asymptotic theory of period spacing for $g$-modes, there is an equation of
\begin{equation}
\bar{\triangle \texttt{P}(l)}=\frac{2\pi^{2}}{\sqrt{l(l+1)}{\int_{0}}^{R}\frac{N}{r}dr},
\end{equation}
\noindent where $N$ is the Brunt-V\"ais\"al\"a frequency and $R$ is the stellar radius. According to Eq.\,(3), neighbouring $g$-modes with the same spherical harmonic degree will show constant period spacing.

It can be noticed in Table 1 that sometimes too many observed pulsation modes crowd in a small period interval, so that they could not be identified as eigenmodes with the same $l$. In particular, we notice that $f_{26}$ is located in the middle of $f_{25}$ and $f_{27}$ when sorting by period. Because $f_{25}$ and $f_{27}$ have already been identified as components of a triplet, then $f_{26}$ can only be identified as an $l$=2 mode. However, only one component has been observed in this case, and we hypothesize it as $m=0$ one. This approximation will be applied to all other modes having been observed as the only component of the corresponding multiplet.

Besides, $f_{2}$ and $f_{3}$ are too much close to each other, and they out to have different spherical harmonic degrees. The spherical harmonic degree $l$ indicates the number of nodal lines of a pulsation mode, which segment the whole stellar surface into several parts. Usually two neighbouring parts move in opposite directions, leading to one part becoming brighter while the other part becoming fainter. As a result, the bigger the spherical harmonic degree, the smaller the amplitude of a pulsation mode due to cancelation of neighbouring parts. According to result of observations in 2008, $f_{2}$ showed an amplitude of 2.1\,mma, while $f_{3}$ showed an amplitude of 12.7\,mma which was much larger than that of $f_{2}$. Therefore, we assume $f_{2}$ as an $l$=2 mode and $f_{3}$ as an $l$=1 mode. It can also be noticed that $f_{3}$ appeared in results of all observations, while $f_{2}$ only appeared in result of observations in 2008.

Similarly, $f_{5}$, $f_{6}$ and $f_{7}$, which was previously identified as a triplet by Provencal et al. (2012), are too much close to each other, so that it is much more difficult to identify their spherical harmonic degree. Since $f_{5}$ and $f_{8}$ have already been identified as two components of a quintuplet, therefore $f_{5}$ is an $l=2$ mode. It can be noticed in Table 1 that $f_{6}$ always shows the largest amplitude among the three modes, and can then be identified according to above argument as an $l=1$ mode. But what should $f_{7}$ be identified as ? One possibility is to identify $f_{7}$ as an $l=3$ mode. It can be noticed in Table 1 that the amplitude of $f_{7}$ is the smallest among the three considered modes, which might support this hypothesis. Another possibility is to identify it as an $m\neq0$ component of a multiplet. Similar situation happens to $f_{24}$, whose period is close to those of $f_{22}$ and $f_{23}$. As $f_{22}$ has already been identified as a member of a quintuplet and $f_{23}$ as a member of a triplet, $f_{24}$ can only be identified as an $l$=3 mode. Otherwise it can also be an $m\neq0$ component of an unidentified multiplet. In addition, $f_{18}$ is only 9.728\,s smaller than $f_{19}$ ($l$=2,$m$=0) and 16.820\,s smaller than $f_{21}$ ($l$=1,$m$=0). Therefore, $f_{18}$ may also be an $l$=3 mode because of too small period differences to $f_{19}$ and $f_{21}$. Otherwise it can be an $m\neq0$ component of an unidentified multiplet. The amplitude of $f_{18}$ is one of the smallest in Table 1, which might support the hypothesis.

It is worth to note that by introducing three $l$=3 modes, we solve the frequency splitting problem mentioned in the introduction. It should be noticed also that the amplitude of $f_{24}$ might not be small as seen in Table 1. Thompson et al. (2008) studied a DAV star named G29-38 with time series of optical spectroscopy and found that a mode was possible to be $l$=4 or 3 by studying limb darkening effect. The amplitude of that mode was 10.7\,mma, which was the fourth-largest among 8 independent modes. Therefore, it is possible that $f_{7}$, $f_{24}$, and $f_{18}$ belong to $l$=3 modes for EC14012-1446.

After above analyses, we have already identified 6 $l$=1 modes, 4 $l=2$ modes, and 3 $l=3$ modes. There are still 12 modes remaining to be identified later. We show these 25 eigenmodes in Table 3, as well as the period difference between two identified modes. It can be noticed that these 4 modes with $l$=2 are likely to show an average period spacing of about 26\,s. According to Eq.\,(3), $\bar{\triangle \texttt{P}(1)}$ : $\bar{\triangle \texttt{P}(2)}$ : $\bar{\triangle \texttt{P}(3)}$ = $\sqrt{6}$ : $\sqrt{2}$ : 1. Consequently we may estimate an uniform period spacing of 45\,s for $l=1$ modes and an uniform period spacing of 18\,s for $l$=3 modes. The period difference of 66.401\,s is obviously larger than 45\,s for $l$=1 modes and the period difference of 24.152\,s is larger than 18\,s for $l$=3 modes. We suggest that there must be strong mode trapping effect on the star, which results in some period differences apart from corresponding uniform period spacing.

The period difference between $f_{29}$ and $f_{30}$ is 17.590\,s, which is consistent with $\bar{\triangle \texttt{P}(3)}$. In addition, $f_{29}$ is only appeared in result of observations in 2008 with Ampl.=1.9\,mma and $f_{30}$ is only appeared in result of observations in April 2004 with Ampl.=2.9\,mma. Therefore, they may be two $l$=3 modes. However, we can not go a step further to identify the rest 10 modes shown in Table 3. We will fit them by modes with either $l$=1 or $l$=2 in the following work.

\section{Model fittings on EC14012-1446}

 \subsection{Input physics and model calculations}

In this section, we discuss our input physics and model calculations. White dwarf models are generated by WDEC, which was first developed by Schwarzschild and subsequently improved by Kutter \& Savedoff (1969), Lamb \& van Horn (1975), and Wood (1990). Itoh et al. (1983) reported radiative opacities and conductive opacities. Lamb (1974) contributed to equation of state in degenerate and ionized core and Saumon, Chabrier \& Van Horn (1995) contributed to equation of state in radiative and thin envelope. The mixing length theory is from B\"{o}hm \& Cassinelli (1971) and Tassoul et al. (1990). The mixing length parameter is adopted as 0.6, which is the same as Bergeron et al. (1995) did.

\begin{table}
\begin{center}
\begin{tabular}{lccccc}
\hline
MS           &WD(MESA)     &WD(WDEC)     \\
($M_{\odot}$)&($M_{\odot}$)&($M_{\odot}$)\\
\hline
1.5          &0.572        &0.560-0.575  \\
2.0          &0.580        &0.580-0.595  \\
2.8          &0.614        &0.600-0.620  \\
3.0          &0.633        &0.625-0.645  \\
3.2          &0.659        &0.650-0.670  \\
3.4          &0.689        &0.675-0.695  \\
3.5          &0.704        &0.700-0.715  \\
3.6          &0.723        &0.720-0.735  \\
3.8          &0.751        &0.740-0.765  \\
4.0          &0.782        &0.770-0.785  \\
4.5          &0.805        &0.790-0.820  \\
\hline
\end{tabular}
\caption{Masses of main-sequence progenitors with corresponding white dwarfs in MESA and connecting white dwarfs in WDEC.}
\end{center}
\end{table}

The core compositions of white dwarf models are results of nuclear burning processes before corresponding stars evolve as white dwarfs. We improved our treatment by adopting the core composition profiles that are directly from evolutionary white dwarf models in a file named make\_co\_wd from MESA version 4298 (Paxton et al. 2011). We show in Table 4 masses of main-sequence progenitors with corresponding white dwarfs in MESA and connecting white dwarfs in WDEC. MESA takes thermonuclear reaction rates of Caughlan \& Fowler (1988) and Angulo et al. (1999). Instead of doing linear fittings to carbon profile as Chen \& Li (2014), we insert MESA core composition profiles (carbon profile, oxygen profile = 1 - carbon profile) directly into WDEC with corresponding structural parameters (mass, radius, luminosity, pressure, temperature, and entropy). Taking the scheme of element diffusion developed by Thoul, Bahcall, \& Loed (1994), Su et al. (2014) added H, He, C, and O diffusion into WDEC. We treat the element diffusion processes according to Su et al. (2014), rather than simply using equilibrium profiles for transition zones of C/He and He/H.

A four-parameter space is made including total stellar mass ($M_{*}$), effective temperature ($T_{\rm eff}$), helium layer mass ($M_{\rm He}$), and hydrogen layer mass ($M_{\rm H}$). The grid of $M_{*}$ is from 0.560\,$M_{\odot}$ to 0.820\,$M_{\odot}$ with a step of 0.005\,$M_{\odot}$. The grid of $T_{\rm eff}$ is from 10800\,K to 12400\,K with a step of 50\,K. Log($M_{\rm He}/M_{*}$) is from -4.0 to -2.0 with a step of 0.5, and log($M_{\rm H}/M_{*}$) is from -10.0 to -4.0 with a step of 0.5. We make more than 100,000 white dwarf models. Then we numerically solve for each model full equations of linear and adiabatic oscillation, finding each eigenmode by scanning in period. Such a fine mesh of white dwarf models has been used to fit EC14012-1446.

 \subsection{Selection of the best-fitting model}

\begin{figure}
\begin{center}
\includegraphics[width=8.2cm,angle=0]{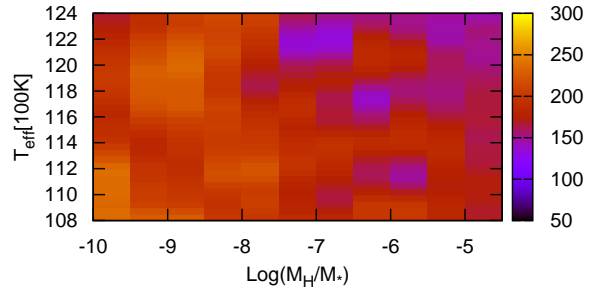}
\end{center}
\caption{Color diagram of $n\phi$. The ordinate is $T_{\rm eff}$ and the abscissa is log($M_{\rm H}/M_{*}$). The diagram shows different values of $n\phi$ by different colors with $M_{*}$=0.710\,$M_{\odot}$ and log($M_{\rm He}/M_{*}$)=-2.5.}
\end{figure}

\begin{table*}
\begin{center}
\begin{tabular}{lllllllllllll}
\hline
$P_{\rm mod}(l,k)$&$P_{\rm obs}$&$\sigma_{p}$&$P_{\rm mod}(l,k)$&$P_{\rm obs}$&$\sigma_{p}$&$P_{\rm mod}(l,k)$&$P_{\rm obs}$&$\sigma_{p}$\\
(s)           &(s)      &(s)         &(s)           &(s)      &(s)         &(s)           &(s)      &(s)         \\
\hline
372.328(1,5)  &         &            &409.434(2,12) &398.715  &$\,$-10.719 &1015.067(2,35)&         &            \\
397.423(1,6)  &399.222  &$\,$ 1.799  &439.167(2,13) &433.887  &$\,$-5.280  &1041.695(2,36)&         &            \\
493.623(1,7)  &         &            &462.357(2,14) &         &            &1066.592(2,37)&1069.085 &$\,$ 2.493  \\
530.411(1,8)  &529.796  &$\,$-0.605  &483.839(2,15) &         &            &              &         &            \\
544.238(1,9)  &         &            &511.579(2,16) &         &            &533.307(3,25) &530.911  &$\,$-2.396  \\
610.947(1,10) &612.125  &$\,$ 1.178  &537.615(2,17) &537.563  &$\,$-0.052  &549.710(3,26) &         &            \\
655.197(1,11) &645.933  &$\,$-9.264  &554.101(2,18) &563.384  &$\,$ 9.283  &568.241(3,27) &         &            \\
680.401(1,12) &678.526  &$\,$-1.875  &581.362(2,19) &         &            &590.332(3,28) &         &            \\
723.869(1,13) &721.855  &$\,$-2.014  &611.159(2,20) &         &            &609.592(3,29) &         &            \\
769.857(1,14) &769.343  &$\,$-0.514  &628.890(2,21) &         &            &628.195(3,30) &         &            \\
808.980(1,15) &805.540  &$\,$-3.440  &657.606(2,22) &657.214  &$\,$-0.392  &646.899(3,31) &         &            \\
851.549(1,16) &         &            &691.203(2,23) &         &            &666.714(3,32) &         &            \\
897.146(1,17) &         &            &714.333(2,24) &714.763  &$\,$ 0.430  &686.631(3,33) &         &            \\
946.185(1,18) &         &            &740.795(2,25) &         &            &705.783(3,34) &705.035  &$\,$-0.748  \\
992.527(1,19) &         &            &769.927(2,26) &771.766  &$\,$ 1.839  &724.162(3,35) &729.187  &$\,$ 5.025  \\
1037.070(1,20)&         &            &794.539(2,27) &         &            &743.881(3,36) &         &            \\
1078.139(1,21)&         &            &822.658(2,28) &         &            &765.954(3,37) &         &            \\
1128.504(1,22)&         &            &849.236(2,29) &         &            &786.082(3,38) &         &            \\
1178.463(1,23)&         &            &872.384(2,30) &         &            &804.261(3,39) &         &            \\
1212.107(1,24)&1217.448 &$\,$ 5.341  &901.750(2,31) &905.591  &$\,$ 3.841  &824.085(3,40) &         &            \\
              &         &            &932.241(2,32) &         &            &844.673(3,41) &         &            \\
356.978(2,10) &350.121  &$\,$-6.857  &954.586(2,33) &         &            &864.225(3,42) &865.108  &$\,$ 0.883  \\
381.763(2,11) &         &            &981.472(2,34) &979.299  &$\,$-2.173  &883.859(3,43) &882.698  &$\,$-1.161  \\
\hline
\hline
\end{tabular}
\caption{The fitting result of the best-fitting model. $\sigma_{p}$ = $P_{\rm obs}$ - $P_{\rm mod}(l,k)$ in seconds.}
\end{center}
\end{table*}

We try to use calculated frequencies of each grid model to fit 25 observed modes in Table 3, among them 6 being identified as $l$=1, 4 as $l$=2, 5 $l$=3, and 10 being possibly $l$=1 or 2. When doing model fittings, we introduce a judging equation,
\begin{equation}
\phi=\frac{1}{n}\sum(|P_{\rm obs}(l,k)-P_{\rm mod}(l,k)|).
\end{equation}
\noindent In Eq.\,(4), $P_{\rm obs}(l,k)$ is the observed period, $P_{\rm mod}(l,k)$ the calculated period, and $n$ the total number of observed modes. A model with the minimum $\Phi$ is selected as the best-fitting one.

After doing model fittings, we find that the fitting results always tend to a large $M_{*}$. It is reasonable and necessary. Only a large $M_{*}$ can result in a small average period spacing, which makes enough modes to fit 25 observed modes in a small range from 350.121\,s to 1217.448\,s. If we take a smaller $M_{*}$, we will obtain a larger average period spacing and the fittings always skip some observed modes. Our best-fitting model has $M_{*}$=0.710\,$M_{\odot}$, log\,$g$=8.261, $T_{\rm eff}$=12200\,K, log($M_{\rm H}/M_{*}$)=-7.0, log($M_{\rm He}/M_{*}$)=-2.5, and $n\phi$=79.613\,s. For calculated $l$= 1, 2, and 3 modes, asymptotic period spacing is 46.870\,s, 27.060\,s, and 19.135\,s, which are close to 45\,s, 26\,s, and 18\,s, respectively.

Figure 1 shows values of $n\phi$ by different colors for grid models with $M_{*}$=0.710\,$M_{\odot}$ and log($M_{\rm He}/M_{*}$)=-2.5. The ordinate is $T_{\rm eff}$ and the abscissa is log($M_{\rm H}/M_{*}$). In the diagram, we can see that the best-fitting model has $T_{\rm eff}$) around 12200\,K and log($M_{\rm H}/M_{*}$) around -7.0 with $n\phi$ around 80\,s. In addition, we also find that $n\phi$ is around 100\,s when $T_{\rm eff}$ is around 11750\,K and log($M_{\rm H}/M_{*}$) around -6.5. We check the model of $T_{\rm eff}$=11750\,K, log($M_{\rm H}/M_{*}$)=-6.5, and $n\phi$=103.940\,s. For the model, $f_{11}$ is fitted by 599.923\,s with an error of 12.202\,s, $f_{16}$ is fitted by 693.502\,s with an error of 14.976\,s, and $f_{25}$ is fitted by 759.838\,s with an error of 9.505\,s. The three $l$=1 modes identified by frequency splitting are badly fitted, therefore, we abandon the model of $T_{\rm eff}$=11750\,K and log($M_{\rm H}/M_{*}$)=-6.5. For other models, $n\phi$ is larger than 130\,s. Therefore, we choose the model of $T_{\rm eff}$=12200\,K and log($M_{\rm H}/M_{*}$)=-7.0 as the best-fitting one.

We list in Table 5 fitting results of the best-fitting model. The calculated periods and observed periods with fitting errors ($\sigma_{p}$) are displayed clearly. According to the best-fitting model, the remaining 10 modes are identified as 3 $l$=1 and 7 $l$=2 modes. Considering those already identified modes altogether, there are 9 $l$=1, 11 $l$=2, and 5 $l$=3 modes. For the best-fitting model, modes identified by frequency splitting are well fitted. The values of $\sigma_{p}$ are basically smaller than 2\,s fitting $m$=0 modes in Table 2.

Stobie et al. (1995) obtained 5 independent frequencies for EC14012-1446, and 4 of them are close to $f_{2}$, $f_{7}$, $f_{10}$, and $f_{22}$. The period of the fifth frequency he has obtained is 937.2\,s, which can be fitted by 932.241\,s (2,32) of the best-fitting model. In addition, Provencal et al. (2012) combined observations spanning 2004-2008 and derived 22 frequencies. It is amazing to notice that some existing modes disappeared while seven new modes emerged. The seven newly emerged modes are periods of 365.220\,s, 750.249\,s, 819.782\,s, 859.710\,s, 879.865\,s, 965.770\,s, and 1036.540\,s, which can be fitted by 372.328\,s (1,5), 740.795\,s (2,25), 822.658\,s (2,28), 851.549\,s (1,16), 872.384\,s (2,30), 954.586\,s (2,33), and 1037.070\,s (1,20) of the best-fitting model, respectively.

 \subsection{Mode trapping effect}

\begin{figure}
\begin{center}
\includegraphics[width=8cm,angle=0]{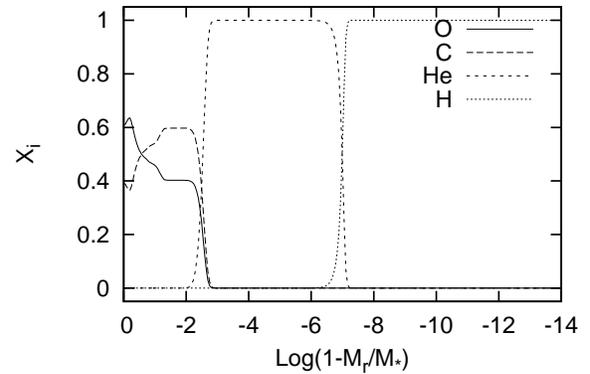}
\end{center}
\caption{Core composition profiles of the best-fitting model.}
\end{figure}

\begin{figure}
\begin{center}
\includegraphics[width=8cm,angle=0]{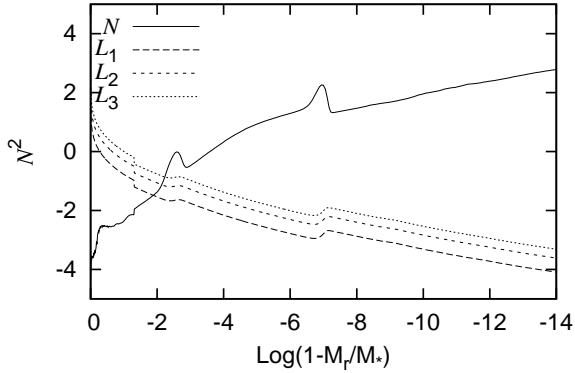}
\end{center}
\caption{Diagram of Brunt-V\"ais\"al\"a frequency $N$ and Lamb frequency $L_{l}$ of the best-fitting model.}
\end{figure}

\begin{figure}
\begin{center}
\includegraphics[width=8cm,angle=0]{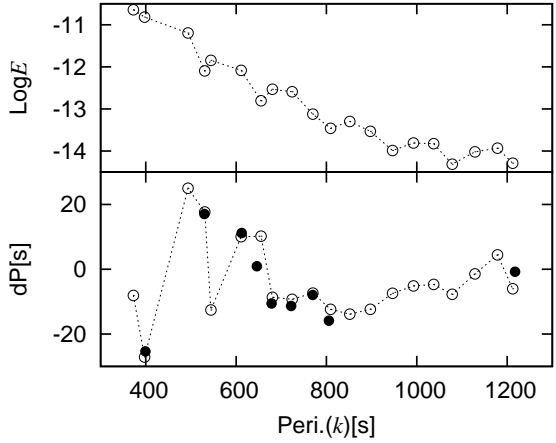}
\end{center}
\caption{Diagram of deviation from the uniform period spacing and mode inertia for $l$=1 modes. The radial orders for open dots are from 5 to 24.}
\end{figure}

\begin{figure}
\begin{center}
\includegraphics[width=8cm,angle=0]{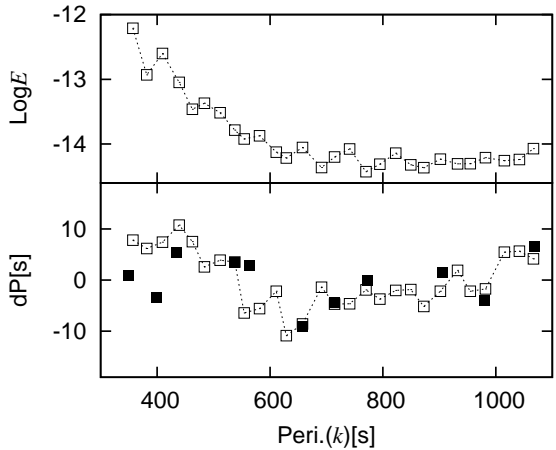}
\end{center}
\caption{Diagram of deviation from the uniform period spacing and mode inertia for $l$=2 modes. The radial orders for open boxes are from 10 to 37.}
\end{figure}

\begin{figure}
\begin{center}
\includegraphics[width=8cm,angle=0]{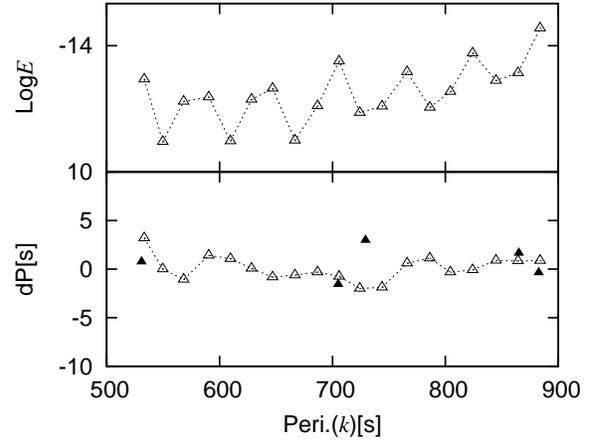}
\end{center}
\caption{Diagram of deviation from the uniform period spacing and mode inertia for $l$=3 modes. The radial orders for open triangles are from 25 to 43.}
\end{figure}

We show in Fig. 2 core composition profiles of the best-fitting model, which are directly from a 3.5\,$M_{\odot}$ main-sequence star, corresponding to a 0.704\,$M_{\odot}$ C/O core, as shown in Table 4. In Fig. 2, the abscissa is set as log(1-$M_{r}/M_{*}$) and the model can be divided clearly into three parts, namely a C/O core, a He layer, and a H envelope. Taking element diffusion into account, profiles of O, C, He, and H are smooth and continuous. In this paper, the best-fitting model has log($M_{\rm H}/M_{*}$)=-7.0 and log($M_{\rm He}/M_{*}$)=-2.5.

In Fig. 3, we show the Brunt-V\"ais\"al\"a frequency and the Lamb frequencies for $l$=1, 2, and 3 of the best-fitting model. It can be seen that there is a small bump at log(1-$M_{r}/M_{*}$)=-2.5 and a big bump at log(1-$M_{r}/M_{*}$)=-7.0, which corresponds to the C/O core to the He layer transition zone and the He layer to the H envelope transition zone, respectively. The composition gradient in the transition zones results in above bumps on the Brunt-V\"ais\"al\"a frequency through the so-called Ledoux term. In addition, there is a smaller bump near the center of the C/O core, which corresponds to oxygen to carbon transition zone near the center of the core.

Winget, Van Horn, and Hansen (1981) first put forward that some modes could be trapped in a specific area, and referred to these modes as trapped modes. A trapped mode means that most of its kinetic energy is restricted in a specific region, which may be confined by composition transition zones. Therefore, the integral in Eq.\,(3) will be carried out not from 0 to $R$ but within a specific area for a trapped mode. For the best-fitting model, if we set the integral range only in the H envelope, the period spacing for modes trapped in the H envelope is 142.337\,s for $l$=1 modes, 82.178\,s for $l$=2 modes, and 58.109\,s for $l$=3 modes, which is three times of the uniform period spacing for $l$=1, 2, and 3 modes, respectively. Actually, the thinner the hydrogen envelope masses, the smaller the integral ranges in Eq.\,(3) for modes trapped in H envelope, and the larger the period spacing of trapped modes.

With the radial orders shown in Table 5, we can calculate the average period spacing for observed modes. They are respectively 44.093\,s, 26.420\,s, and 19.606\,s for $l$=1, 2, and 3 modes. The deviation from uniform period spacing (dP) diagram can be made by
\begin{equation}
dP_{k,l} = P_{k,l} - P_{0} - k*\bar{\triangle \texttt{P}(l)}.
\end{equation}
\noindent In Eq.\,(5), $P_{0}$ is a suitable chosen reference such that $dP_{k,l}$ is close to 0. In addition, mode inertia diagram is usually used to study mode trapping effect. Mode inertia ($E$) is expressed by
\begin{equation}
\ E = \frac{{4\pi\int_{0}^{R}[(|\tilde{\xi}_{r}(r)|^2+l(l+1)|\tilde{\xi}_{h}(r)|^2)]\rho_{0}r^2dr}}{{M_{*}[|\tilde{\xi}_{r}(R)|^2+l(l+1)|\tilde{\xi}_{h}(R)|^2)]}}.
\end{equation}
\noindent In Eq.\,(6), $\rho_{0}$ is the local density, $\tilde{\xi}_{r}(r)$ the radial displacement, and $\tilde{\xi}_{h}(r)$ the horizontal displacement. Mode inertia represents kinetic energy normalized on the surface. The smaller the mode inertia of a mode, the larger the kinetic energy confined on the surface. Therefore, we can select modes trapped in H envelope by choosing the smallest inertia modes.

In Fig. 4, we show deviation from the uniform period spacing and mode inertia diagram for $l$=1 modes. For open dots, the radial order is from 5 to 24. The filled dots are observed modes. In the low panel, observed modes are well fitted by open dots, especially for the 6 $l$=1 modes in Table 3. In the up panel, we can clearly choose 6 modes trapped in H envelope, the radial order of which is $k$=8, 11, 15, 18, 21, and 24, respectively. There are usually two normal modes between two trapped modes. Namely, it is clear that the period spacing for trapped modes is three times of the uniform period spacing. Comparing two panels, for the observed modes, we can find that $f_{6}$, $f_{13}$, $f_{28}$, and $f_{34}$ are trapped in H envelope.

In Fig. 5 and Fig. 6, we show deviation from the uniform period spacing and mode inertia diagram for $l$=2 modes and 3 modes respectively. Open boxes and filled boxes are calculated modes and observed modes for $l$=2. Open triangles and filled triangles are calculated modes and observed modes for $l$=3. In Fig. 5 and Fig. 6, we can also find the phenomena that the period spacing of trapped modes is three times of corresponding uniform period spacing. For observed $l$=2 modes, $f_{14}$ and $f_{26}$ are trapped in H envelope. For observed $l$=3 modes, $f_{24}$ is trapped in H envelope.

According to the asymptotic period spacing law, pulsation periods of different modes with the same $l$ should have an uniform period spacing. However, trapped modes have their own period spacing different from that of normal modes. When ranging all the eigenmodes with the same $l$ together, trapped modes will have relatively small period spacings. For $l$=1 modes, $f_{13}$, which is between $f_{11}$ and $f_{16}$, is trapped in H envelope. It results in the period difference between $f_{11}$ and $f_{16}$ smaller than twice of the uniform period spacing. For $l$=2 modes, $f_{26}$ ($k$=26) is trapped in H envelope, which results in a small period spacing between modes of $k$=26 and $k$=27 and a large period difference between modes of $k$=24 ($f_{22}$) and $k$=26 ($f_{26}$). For $l$=3 modes, $f_{24}$ ($k$=35) is trapped in H envelope and jumps the queue of the uniform period spacing, which results in $f_{24}$ being close to the mode of $k$=36 and far away from the mode of $k$=34 ($f_{18}$). This is why there are period differences around 66\,s for $l$=1 modes, around 57\,s for $l$=2 modes, and around 24\,s for $l$=3 modes in Table 3, which are apart from corresponding uniform period spacing.

\section{Comparisons between the best-fitting model and the previous work}

Taking homogeneous core compositions, namely C/O = 50 : 50, Castanheira \& Kepler (2009) did asteroseismology work on EC14012-1446. They obtained two best-fitting models, model1 and model2. We show the observed periods they adopt and their two best-fitting results in Table 6. They try to fit 9 modes with $\phi$=7.47\,s for model1 and 7.01\,s for model2. Romero et al. (2012) did asteroseismology work on EC14012-1446, adopting fully evolutionary white dwarf models with time-dependent element diffusion. They also fit the same 9 modes and obtain $\phi$=2.54\,s. The best-fitting results are shown in the fourth column in Table 6. In the present paper, we review the observations by Handler et al. (2008), Provencal et al. (2012) and try to solve the frequency splitting question. There are 6 $l$=1, 4 $l$=2, 5 $l$=3, and 10 $l$ = 1 or 2 modes, total 25 eigenmodes identified. The value $\phi$ is 3.185\,s for the best-fitting model. It worth to say that some mode identifications in Table 6 are not the same with which in Table 3.

\begin{table}
\begin{center}
\begin{tabular}{lllllllllll}
\hline
$P_{\rm obs}$&$P_{mod1}(l,k)$&$P_{mod2}(l,k)$&$P_{mod3}(l,k)$ \\
(s)      &(s)            &(s)            &(s)             \\
\hline
398.9    &  403.6(1,5)   &  417.5(1,7)   &403.823(1,7)    \\
530.1    &  532.9(1,7)   &  529.3(1,10)  &524.782(1,10)   \\
610.4    &  610.2(1,9)   &  610.6(1,12)  &613.677(1,12)   \\
678.6    &  663.1(1,10)  &  687.5(1,14)  &675.620(1,14)   \\
722.9    &  722.2(1,11)  &  727.8(1,15)  &721.733(1,15)   \\
769.1    &  770.3(1,13)  &  769.1(1,16)  &769.121(1,16)   \\
882.7    &  862.4(1,14)  &  888.9(1,19)  &883.878(2,34)   \\
937.2    &  916.6(1,15)  &  927.1(1,20)  &934.485(2,36)   \\
1217.4   &  1216.2(1,21) &  1204.0(1,27) &1216.141(1,27)  \\
\hline
\end{tabular}
\caption{Results of Castanheira \& Kepler (2009) and Romero et al. (2012). The first column is observed periods. The second column and the third column are two best-fitting results of Castanheira \& Kepler (2009). The fourth column is the best-fitting results of Romero et al. (2012).}
\end{center}
\end{table}

\begin{table}
\begin{center}
\begin{tabular}{lccccc}
\hline
ID&$T_{\rm eff}$&log\,$g$&$M_{*}$      &log($M_{\rm H}/M_{*}$)&log($M_{\rm He}/M_{*}$)\\
 &(K)      &      &($M_{\odot}$)&                  &                   \\
\hline
1&11900    &8.16  &0.70         &                  &                   \\
2&11768    &8.080 &             &                  &                   \\
3&11600    &      &0.64         &-7.5              &-2.5               \\
4&11500    &      &0.76         &-5.0              &-2.5               \\
5&11709    &8.05  &0.632        &-4.29             &-1.76              \\
6&12200    &8.261 &0.710        &-7.0              &-2.5               \\
\hline
\end{tabular}
\caption{Diagram of best-fitting models. The ID number 1, 2, 3, 4, 5, and 6 means results of Bergeron et al. (2004), Koester et al. (2009), model1 and model2 from Castanheira \& Kepler (2009), Romero et al. (2012), and the present paper, respectively.}
\end{center}
\end{table}

In order to compare the best-fitting results with each other, we set them together in Table 7. The ID number 1, 2, 3, 4, 5, and 6 means results of Bergeron et al. (2004), Koester et al. (2009), model1 and model2 from Castanheira \& Kepler (2009), Romero et al. (2012), and the present paper, respectively. We can see that $T_{\rm eff}$=12200\,K is close to 11900\,K and a little higher than other work. Log($M_{\rm H}/M_{*}$)=-7.0 is close to -7.5 and thinner than -5.0 and -4.29. Log($M_{\rm He}/M_{*}$)=-2.5 is the same with results of Castanheira \& Kepler (2009) and close to the result of Romero et al. (2012). The stellar mass $M_{*}$=0.710\,$M_{\odot}$ is close to 0.70\,$M_{\odot}$ and in the middle of 0.632\,$M_{\odot}$ and 0.76\,$M_{\odot}$.

\section{Discussion and Conclusions}

In the paper, we review observations on EC14012-1446 in April 2004, June 2004, May 2005, April 2007 (Handler er al. 2008), and 2008 (Provencal et al. 2012). There are 34 independent frequencies altogether. Handler et al. (2008) derived a fairly short rotation period about 0.61\,d according to frequency splitting around 9.55\,$\mu$Hz. Provencal et al. (2012) derived a rotation period around 1.53\,d with an averaged frequency splitting of 3.79\,$\mu$Hz. A DAV star has evolved through a very long time therefore the differential rotation effect should be weak. A 1.53\,d/0.61\,d=2.51 times of differential rotation seems impossible. If 3.79\,$\mu$Hz corresponds to triplets and 9.55\,$\mu$Hz corresponds to quintuplets, both 3.79/9.55=0.397 and 3.79/(9.55/2)=0.794 will not be consistent with the theory value of 0.60 by Winget et al. (1991). In addition, we notice that the frequency splitting for $f_{19}$ and $f_{22}$ is 16.32\,$\mu$Hz, for $f_{5}$ and $f_{8}$ is 30.89\,$\mu$Hz. Both 9.55/16.32=0.585 and 9.55/(30.89/2)=0.618 are close to 0.60. In addition, the frequency splitting around 9.55\,$\mu$Hz was appeared many times in the result of Handler er al. (2008). Two modes with frequency splitting of 9.892\,$\mu$Hz were also appeared in the result of Provencal et al. (2012). Therefore, we set modes of frequency splitting around 9.55\,$\mu$Hz as triplets and modes of frequency splitting of 16.32\,$\mu$Hz and 30.89\,$\mu$Hz as quintuplets. For frequency splitting around 3.79\,$\mu$Hz, we introduce $l$=2 and 3 modes. At last, there are 6 $l$=1, 4 $l$=2, 5 $l$=3, and 10 $l$=1 or 2 modes, total 25 eigenmodes identified.

Grids of white dwarf models are generated by WDEC. The core composition profiles are from white dwarf models generated by MESA. The main-sequence stars are from 1.5\,$M_{\odot}$ to 4.5\,$M_{\odot}$ and corresponding white dwarfs are from 0.572\,$M_{\odot}$ to 0.805\,$M_{\odot}$. When white dwarfs cool down by WDEC, element diffusion among H, He, C, and O are adopted. We make a four-parameter space, including $M_{*}$, $T_{\rm eff}$, log($M_{\rm He}/M_{*}$), and log($M_{\rm H}/M_{*}$). When fitting 25 eigenmodes by all the grid models, a best-fitting model is selected, which has $M_{*}$=0.710\,$M_{\odot}$, $T_{\rm eff}$=12200 K, log($M_{\rm He}/M_{*}$)=-2.5, log($M_{\rm H}/M_{*}$)=-7.0, log\,$g$=8.261, and $n\phi$=79.613\,s namely $\phi$=3.185\,s. For the best-fitting model, modes of frequency splitting are well fitted. According to the best-fitting model, the 10 $l$=1 or 2 modes are identified as 3 $l$=1 and 7 $l$=2. The average period spacing is respectively 44.093\,s, 26.420\,s, and 19.606\,s for observed $l$=1, 2, and 3 modes.

We discuss the mode trapping effect for EC14012-1446. The H envelope is thin for log($M_{\rm H}/M_{*}$)=-7.0. According to the asymptotic period spacing law, the period spacing for modes trapped in H layer is three times of the uniform period spacing. The mode inertia diagram is used to research on the mode trapping effect. The trapped modes are very clear on the mode inertia diagram. Between two trapped modes, there are usually two normal modes. For observed $l$=1 modes, $f_{6}$, $f_{13}$, $f_{28}$, and $f_{34}$ are trapped modes. For observed $l$=2 modes, $f_{14}$ and $f_{26}$ are trapped modes. For observed $l$=3 modes, $f_{24}$ is a trapped mode.

\section{Acknowledgements}
This work is supported by the Knowledge Innovation Key Program of the University of Chinese Academy of Sciences under Grant No. KJCX2-YW-T24 and the Foundation of XDB09010202. We are very grateful to J. N. Fu, C. Li, and J. Su for their kindly discussion and suggestions.

\label{lastpage}

\end{document}